\documentclass{article}
\usepackage{graphicx}
\usepackage[english]{babel}
\usepackage{psfrag}

\def\slashchar#1{\setbox0=\hbox{$#1$}
   \dimen0=\wd0
   \setbox1=\hbox{/} \dimen1=\wd1
   \ifdim\dimen0>\dimen1
      \rlap{\hbox to \dimen0{\hfil/\hfil}}
      #1
   \else
      \rlap{\hbox to \dimen1{\hfil$#1$\hfil}}
      /
   \fi}

\def\bei{\begin{itemize}}
\def\ei{\end{itemize}}

\def\beeq{\begin{eqnarray}} 
\def\beqa{\begin{eqnarray}}
\def\bea{\begin{eqnarray}}

\def\eea{\end{eqnarray}}
\def\eqa{\end{eqnarray}}
\def\eeeq{\end{eqnarray}}

\def\eqar{\end{array}}
\def\beqar{\begin{array}}

\def\beas{\begin{eqnarray*}}
\def\beqas{\begin{eqnarray*}}

\def\eqas{\end{eqnarray*}}
\def\eeas{\end{eqnarray*}}

\def\beq{\begin{equation}} 
\def\be{\begin{equation}}

\def\ee{\end{equation}}
\def\eq{\end{equation}}
\def\eeq{\end{equation}}

\def\beqd{\begin{displaymath}}
\def\eeqd{\end{displaymath}}
\def\eqd{\end{displaymath}}

\def\beeq{\begin{eqnarray}} \def\eeeq{\end{eqnarray}}

\newcommand{\fin}{\end{document}}

\newcommand{\veckone}{{\bf k}_1}
\newcommand{\vecktwo}{{\bf k}_2}

\newcommand{\veckjone}{{\bf k}_{J,1}}
\newcommand{\veckjtwo}{{\bf k}_{J,2}}

\newcommand{\deins}[1]{{\rm d}#1\,}
\newcommand{\dzwei}[1]{{\rm d}^2#1\,}

\newcommand{\dkone}{\dzwei{\veckone}}
\newcommand{\dktwo}{\dzwei{\vecktwo}}

\newcommand{\dsigma}{\deins{\sigma}}
\newcommand{\dsigmahat}{\deins{{\hat\sigma}_{\rm{ab}}}}

\newcommand{\dxone}{\deins{x_1}}
\newcommand{\dxtwo}{\deins{x_2}}
\newcommand{\dyjetone}{\deins{y_{J,1}}}
\newcommand{\dyjettwo}{\deins{y_{J,2}}}

\newcommand{\dphijone}{\deins{\phi_{J,1}}}
\newcommand{\dphijtwo}{\deins{\phi_{J,2}}}
\newcommand{\dtwojets}{{\rm d}|\veckjone|\,{\rm d}|\veckjtwo|\,\dyjetone \dyjettwo}

\newcommand{\shat}{{\hat s}}

\title{
\includegraphics[width=0.35\textwidth]{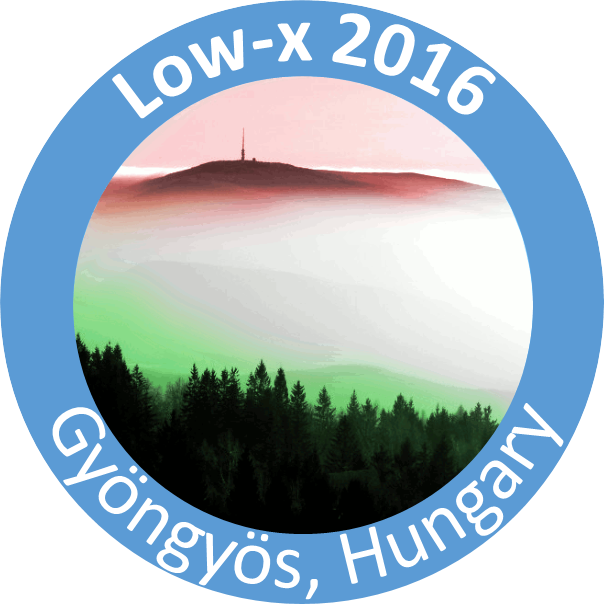}\\[1cm]
      Probing BFKL dynamics in Mueller-Navelet jet production at the LHC}
\author{{B. Duclou\'e$^{1,2}$, L. Szymanowski$^3$, S. Wallon$^{4,5}$}\\[1ex]
$^1$Department of Physics, University of Jyv\"askyl\"a, \\ P.O. Box 35, 40014 University of Jyv\"askyl\"a, Finland\\
$^2$Helsinki Institute of Physics, P.O. Box 64, \\ 00014 University of Helsinki, Finland\\
$^3$National Centre for Nuclear Research, \\ Ho\.za 69, 00-681 Warsaw, Poland \\
$^4$Laboratoire de Physique Th\'eorique, UMR 8627, CNRS, \\ Univ. Paris Sud,  Universit\'e Paris-Saclay, 91405 Orsay, France   \\
$^5$UPMC Univ. Paris 06, Facult\'e de Physique, \\ 4 place Jussieu,  75252 Paris Cedex 05, France\\
}

\begin{document}

\fontfamily{lmss}\selectfont
\maketitle

\begin{abstract}
We review the results of our studies on the production of two  jets with a large interval of rapidity at hadron colliders, which was proposed by Mueller and Navelet as a possible test of the high energy dynamics of QCD, within  the next-to-leading logarithm framework. The application of the Brodsky-Lepage-Mackenzie procedure to fix the renormalization scale leads to a very good description of the available CMS data at the LHC for the azimuthal correlations of the jets. We show 
that the inclusion of next-to-leading order corrections to the jet vertex significantly reduces the importance of energy-momentum non-conservation which is inherent to the BFKL approach, for an asymmetric jet configuration. 
\end{abstract}

\begin{figure}[h]
 \centerline{\includegraphics[height=6cm]{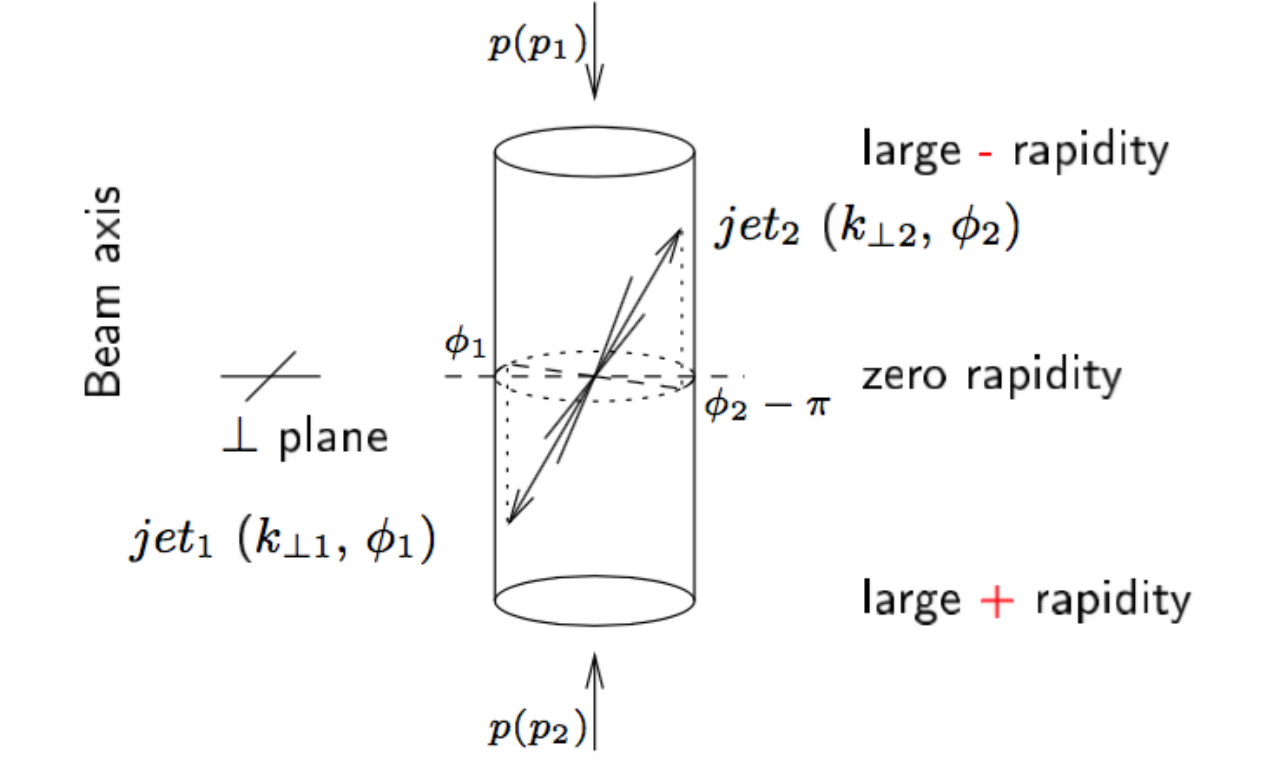}}
 \caption{Mueller Navelet jets production.}
\label{JetsGeneral}
\end{figure}

\vspace*{1cm}
One of the most famous testing grounds for BFKL physics 
\cite{Fadin:1975cbKuraev:1976geKuraev:1977fsBalitsky:1978ic}
are the Mueller Navelet jets \cite{Mueller:1986ey}, illustrated in Fig. \ref{JetsGeneral}.
Besides the cross section also a more exclusive observable within this process drew the attention, namely the azimuthal correlation between these jets. Considering hadron-hadron scattering in the common parton model to describe two jet production at LO, one deals with a back-to-back reaction and expects the azimuthal angles of the two jets always to be $\pi$ and hence completely correlated. This corresponds in Fig. \ref{JetsGeneral} to  $\phi_{J,1}=\phi_{J,2}-\pi$. But when we increase the rapidity difference between these jets, the phase space allows for more and more emissions leading to an angular decorrelation between the jets. 

\begin{figure}[h]
\includegraphics[height=5.2cm,angle=0]{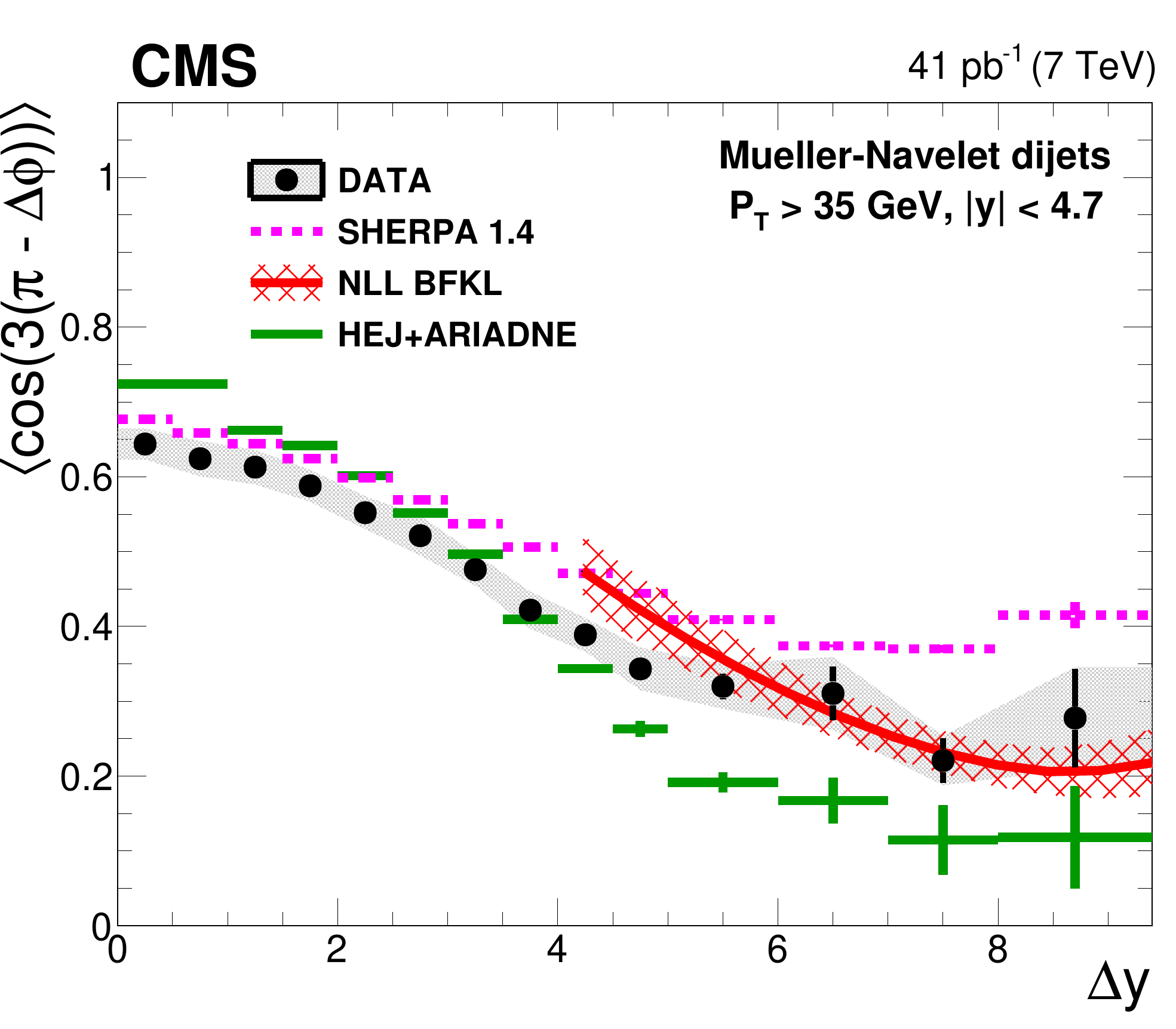}
\includegraphics[height=5.2cm,angle=0]{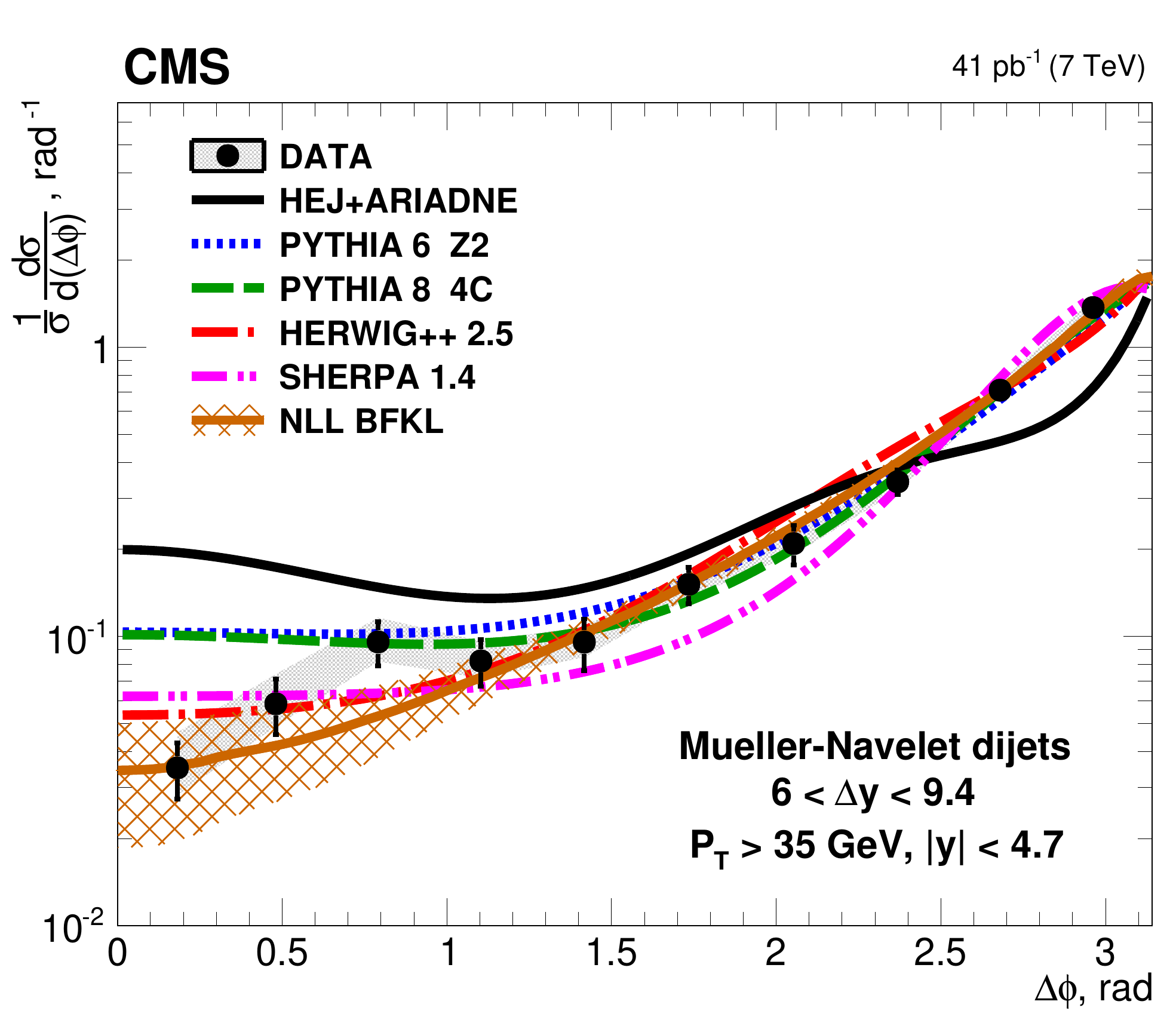}
 \caption{The comparison of theoretical calculation of Ref.  \cite{Ducloue:2013bva} for $\langle \cos 3(\pi - \Delta \phi) \rangle$ (left panel) and $ \frac{1}{{\sigma}}\frac{d{\sigma}}{d \varphi}$ (right panel),
with the measurements by CMS@LHC  presented in \cite{Khachatryan:2016udy}. }
\label{Cos3Phi and angular distribution} 
\end{figure}

\begin{figure}[h]
\includegraphics[height=5.2cm,angle=0]{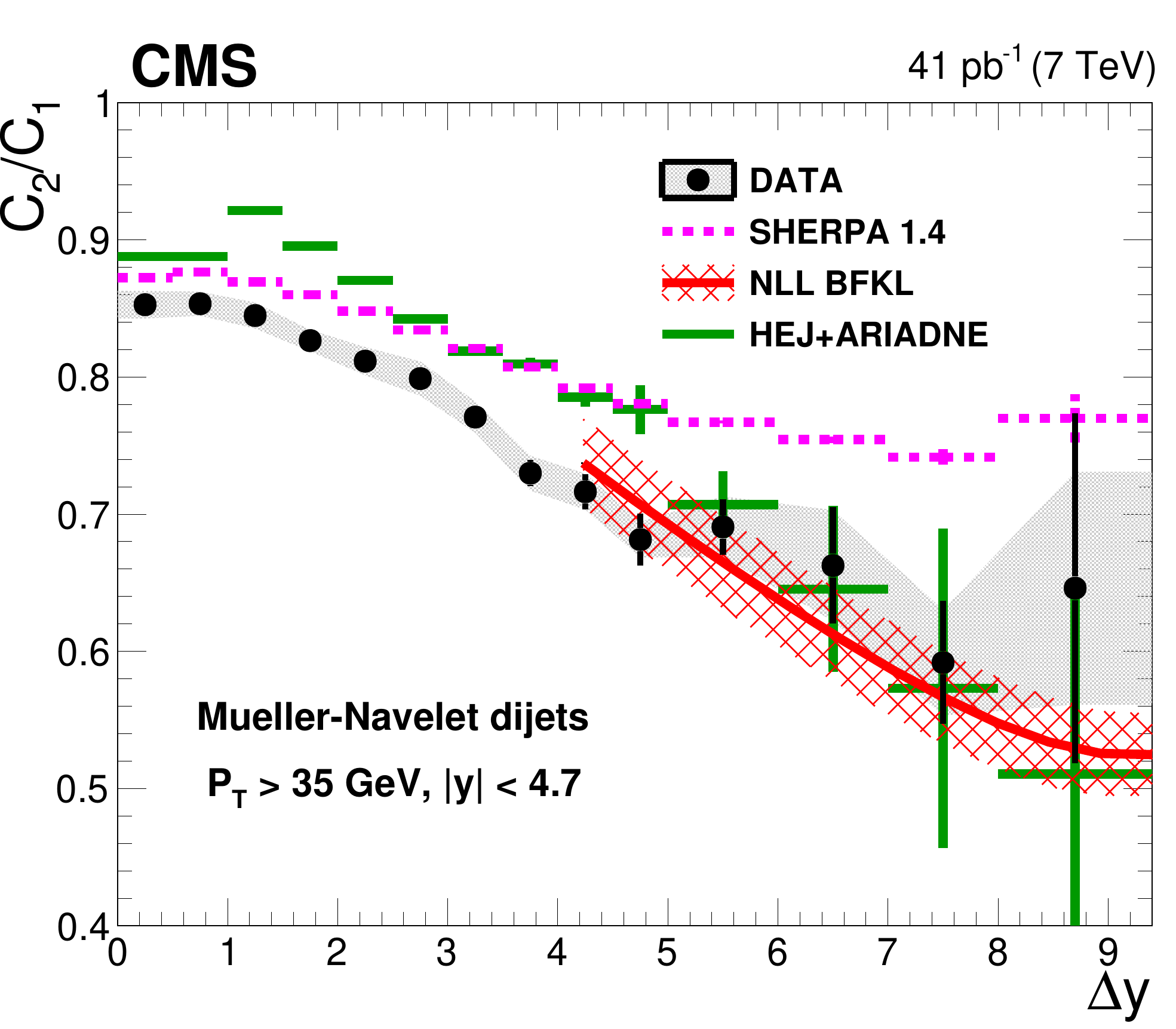}
\includegraphics[height=5.2cm,angle=0]{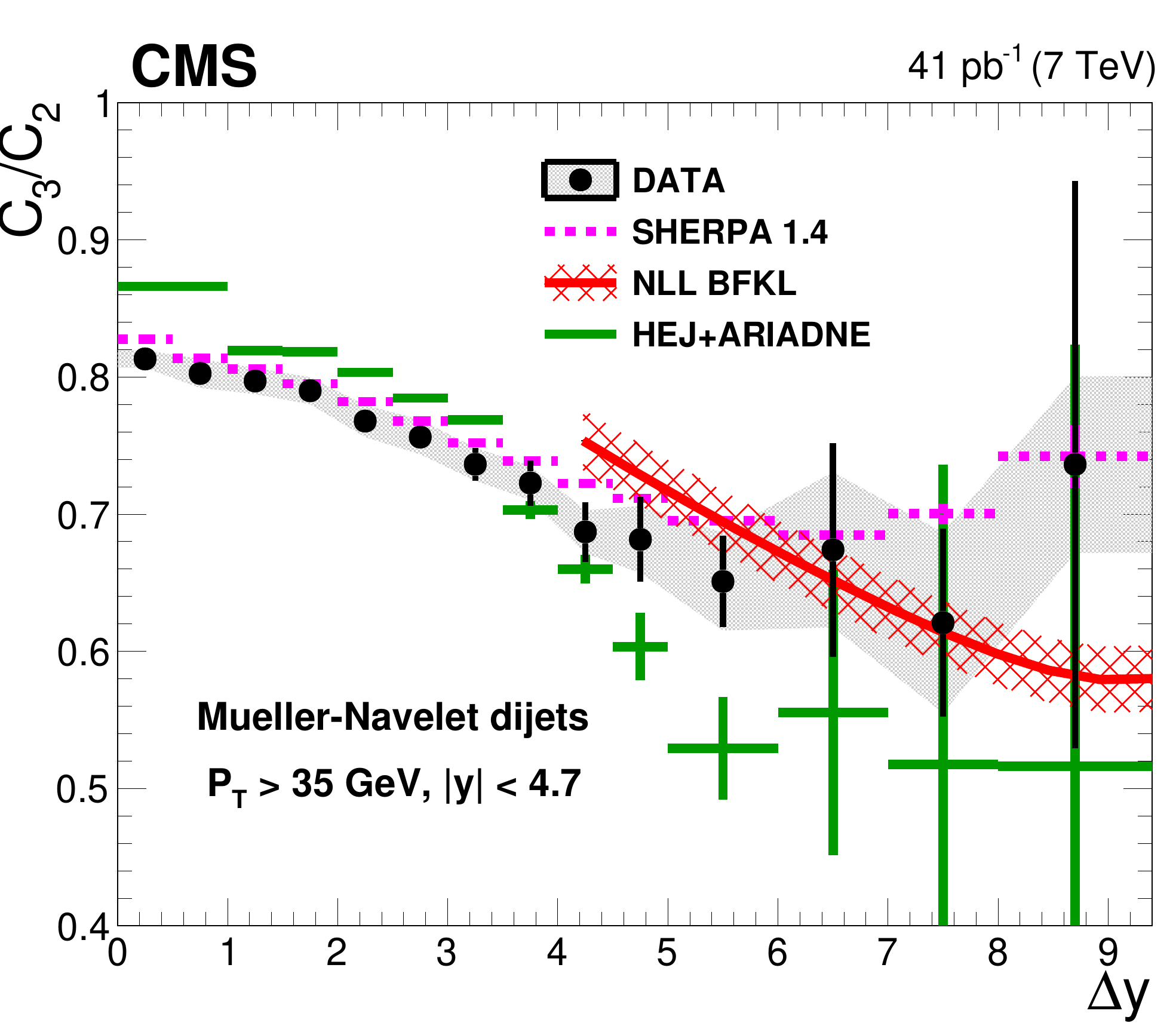}
 \caption{The comparison of theoretical predictions of Ref.  \cite{Ducloue:2013bva} for the ratio $\frac{\cos 2\varphi}{\cos \varphi}$ (left panel) and 
 the ratio $\frac{\cos 3\varphi}{\cos 2\varphi}$ (right panel), with the measurements by CMS@LHC  presented in \cite{Khachatryan:2016udy}.
 }
\label{ratios1/2and2/3 }
\end{figure}

The production of two jets of transverse momenta $\veckjone$, $\veckjtwo$ and rapidities
$y_{J,1}$, $y_{J,2}$ is described by the differential cross-section 
\begin{eqnarray}
\label{collinear}
&&  \frac{\dsigma}{\dtwojets} = 
  \\
&&  \sum_{{\rm a},{\rm b}} \int_0^1 \dxone \int_0^1 \dxtwo f_{\rm a}(x_1) f_{\rm b}(x_2)  \frac{\dsigmahat}{\dtwojets},
\nonumber
\end{eqnarray}
where $f_{\rm a, b}$ are the usual collinear partonic distributions (PDF). In the BFKL framework, the partonic cross-section reads
\beqa
  &&\frac{\dsigmahat}{\dtwojets}  = 
  \nonumber \\
&&  \int \dphijone\dphijtwo\int\dkone\dktwo V_{\rm a}(-\veckone,x_1)\,G(\veckone,\vecktwo,\shat)\,V_{\rm b}(\vecktwo,x_2),\label{eq:bfklpartonic}
\eqa
where $V_{\rm a, b}$ and $G$ are respectively the jet vertices and the BFKL Green's function. 
At present, they are known with the next-to-leading logarithm accuracy  \cite{Fadin:1998py,Ciafaloni:1998gs,Bartels:2001ge,Bartels:2002yj,Caporale:2011cc}.
The cross sections (\ref{collinear}, \ref{eq:bfklpartonic}) are the basic blocks of the calculations presented in 
\cite{Colferai:2010wu,Ducloue:2013hia,Ducloue:2013bva}
of the decorrelation coefficients $\langle \cos m (\pi - \Delta \phi) \rangle$, $\Delta \phi=\phi_{J,1}-\phi_{J,2}$, $m \;\in \; N$, which are observables 
which can be measured at experiments performed at the LHC.  At present the measurements of the CMS collaboration are done for the so called the symmetric configuration 
of produced jets, i.e. jets  in which the lower limit on transverse momentum is the same for both jets.The theoretical estimates obtained in this case for  $\langle \cos m (\pi - \Delta \phi) \rangle$ with the use of the Brodsky-Lepage-Mackenzie  method to fix the renormalization
scale 
\cite{Brodsky:1982gc}, turns out to be in good agreement with the measurement reported recently 
by the CMS collaboration  \cite{Khachatryan:2016udy}.  
\begin{figure}[h]
\includegraphics[height=5.2cm,angle=0]{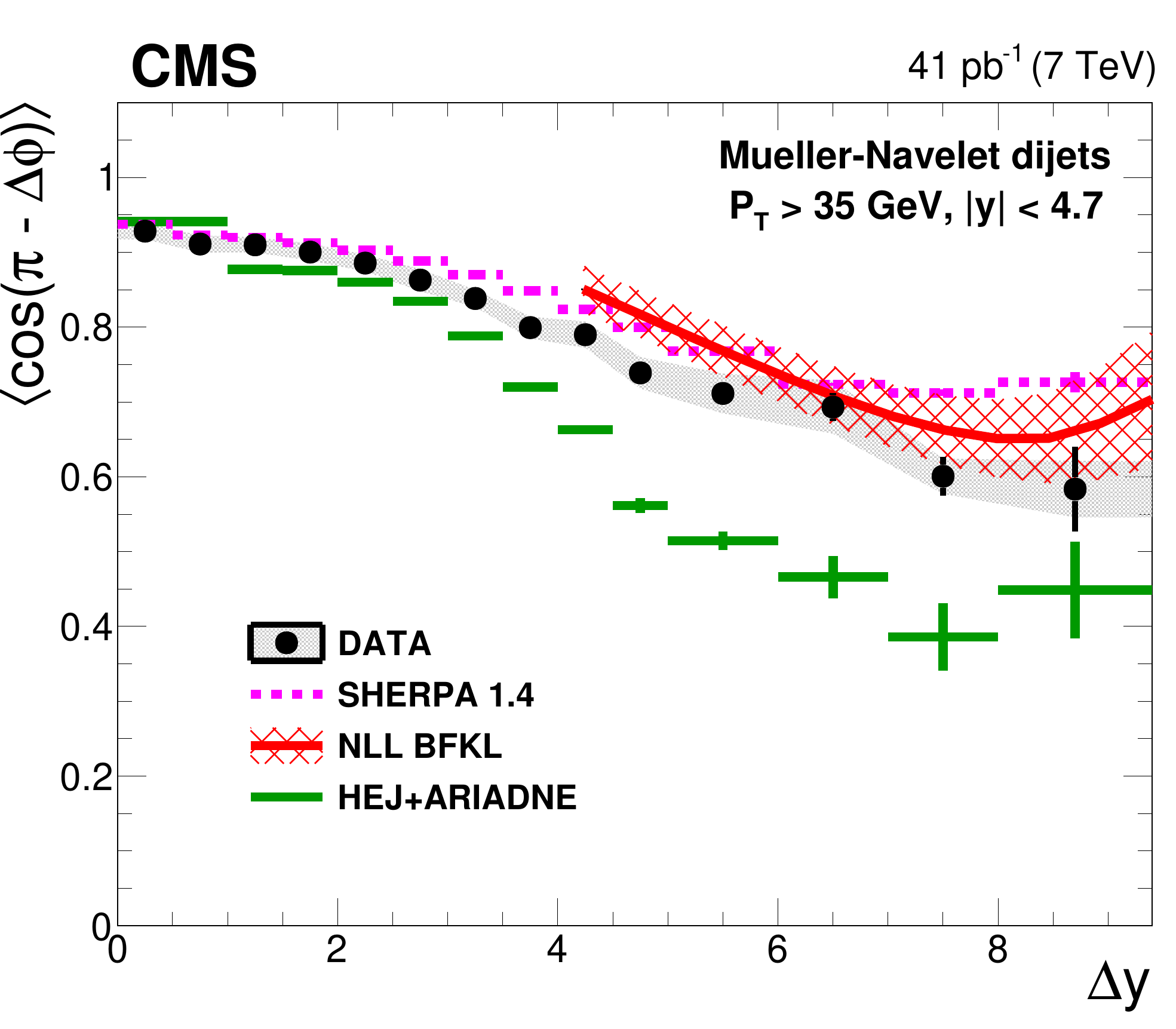}
\includegraphics[height=5.2cm,angle=0]{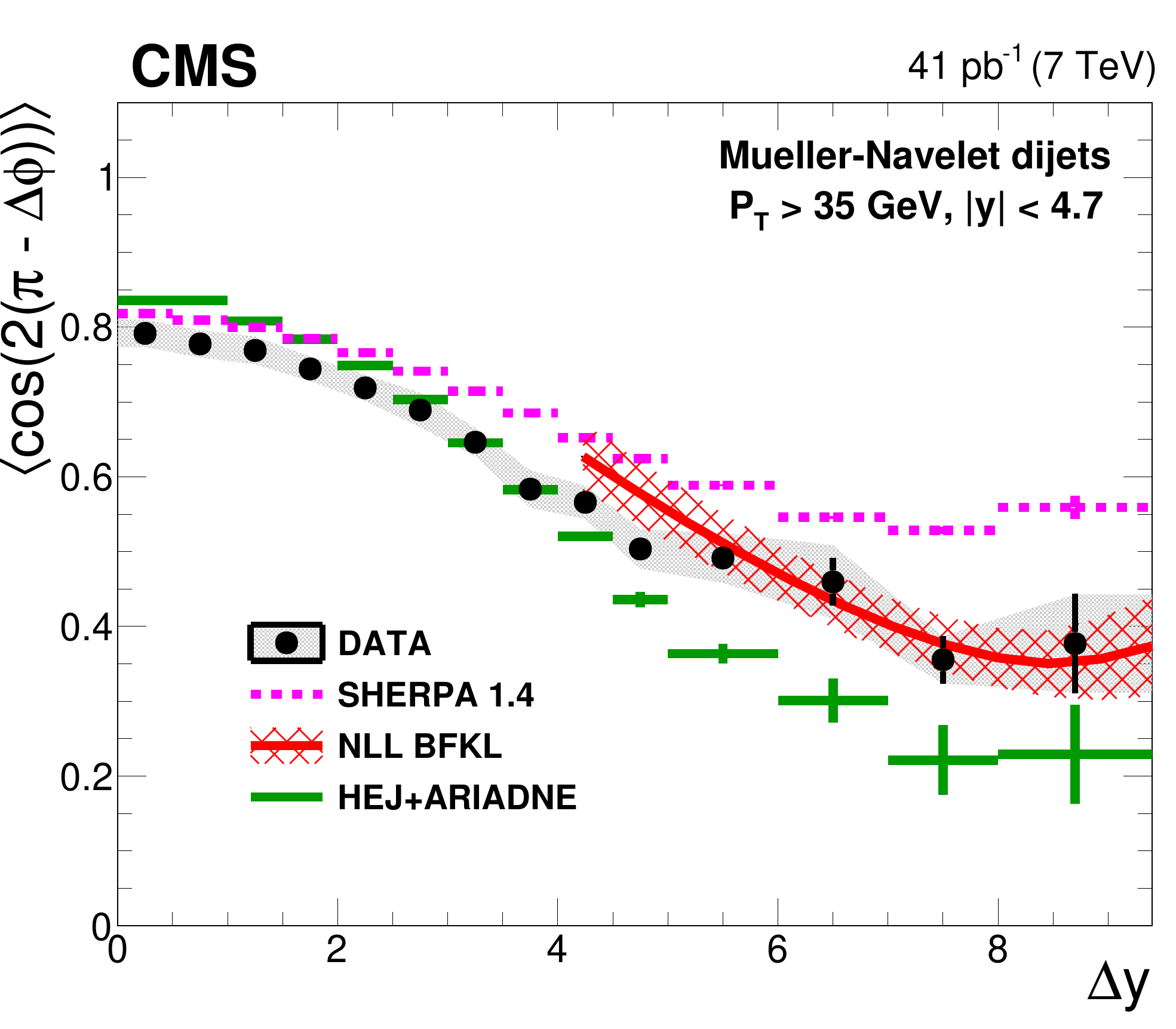}
 \caption{ The comparison of the results of the theoretical calculation of Ref.  \cite{Ducloue:2013bva} for $\langle \cos (\pi - \Delta \phi) \rangle$ (left panel) and $\langle \cos 2 (\pi - \Delta \phi) \rangle$ (right panel),
with the measurements by CMS@LHC  presented in \cite{Khachatryan:2016udy}. }
\label{CosPhiCos2Phi} 
\end{figure}
This fact is clearly illustrated in Fig. \ref{CosPhiCos2Phi} and the left panel of Fig. \ref{Cos3Phi and angular distribution} shown in Ref. \cite{Khachatryan:2016udy}, which also shows  the comparison of measurements with various Monte Carlo simulations. The observables which are more robust against theoretical uncertainties, in particular which are more stable against a choice of renormalization and factorization scales, are the ratios of decorrelation coefficients.
Fig. \ref{ratios1/2and2/3 } shows a good agreeement of results of calculation with the CMS data. 
 The CMS collaboration also measured  
 the azimuthal distribution of the jets, defined as 
\begin{equation}
 \frac{1}{{\sigma}}\frac{d{\sigma}}{d \varphi}
  ~=~ \frac{1}{2\pi}
  \left\{1+2 \sum_{n=1}^\infty \cos{\left(n \varphi\right)}
  \left<\cos{\left( n \varphi \right)}\right>\right\},\;\;\;\;\varphi=\Delta \phi -\pi \,.
\end{equation}
The good agreement between theoretical estimates  of \cite{Ducloue:2013bva} and measurements of this observable is shown in the right panel of Fig. \ref{Cos3Phi and angular distribution}.

\begin{figure}[h]
\vspace*{-0.1cm}
\includegraphics[width=6.4cm]{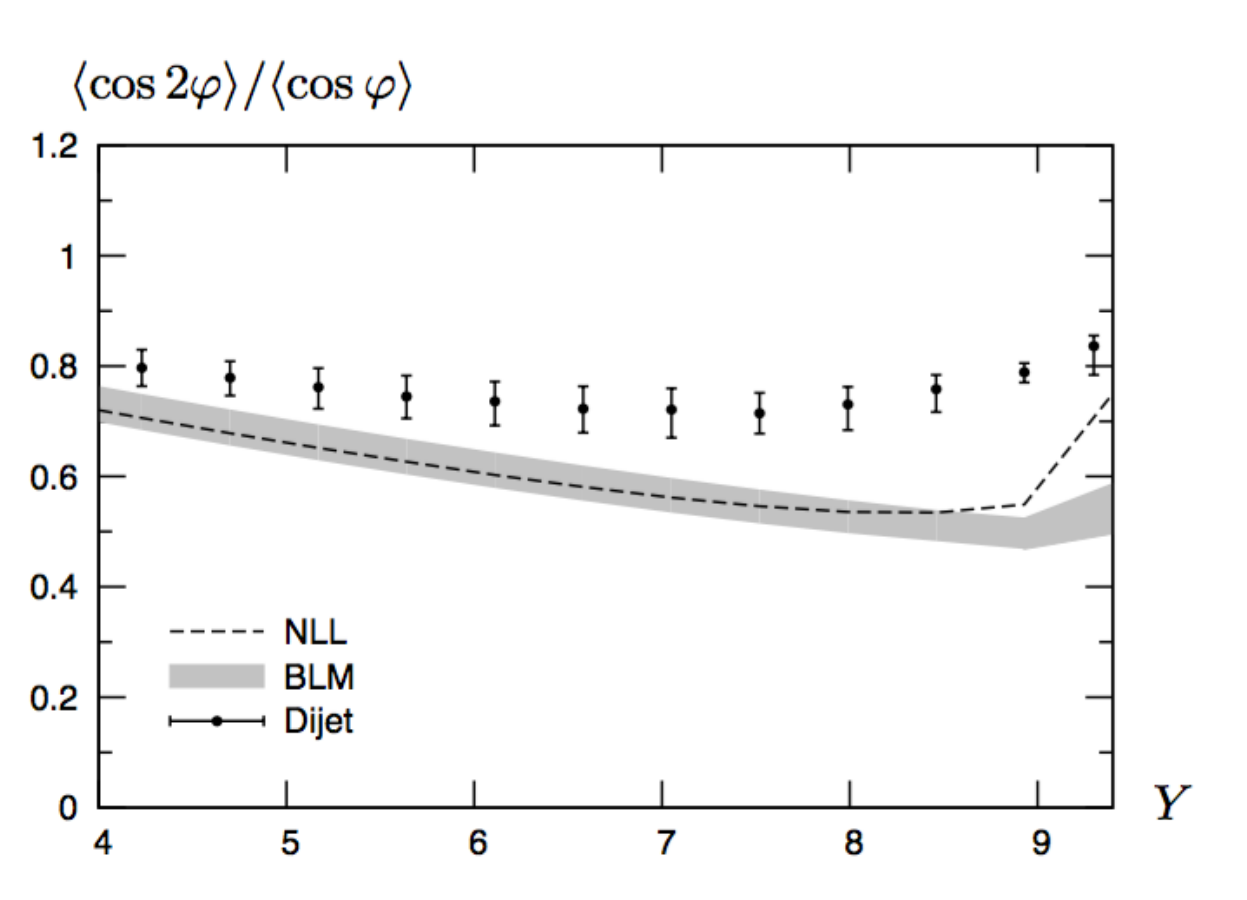} 
\includegraphics[width=7.0cm]{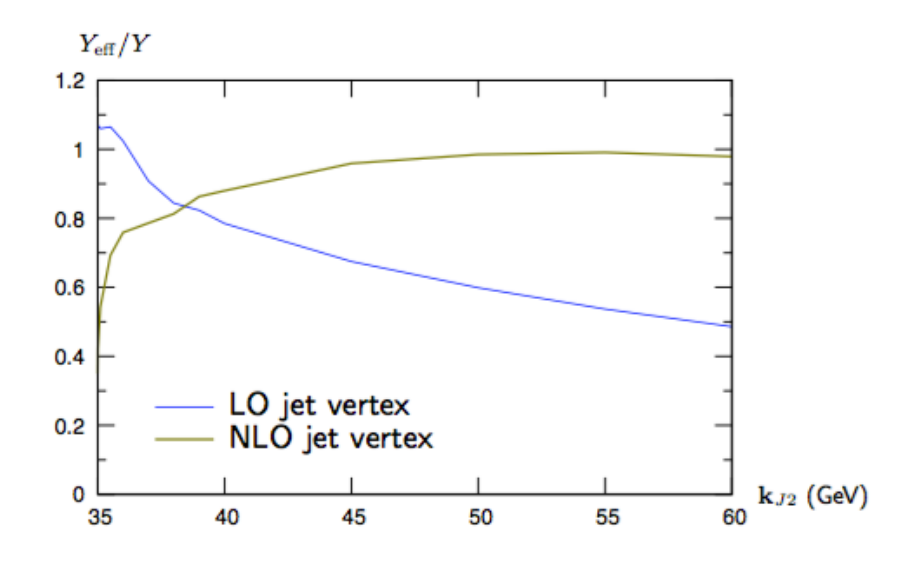}
\caption{Left panel:
Asymmetric configuration. Variation of the ratio  $\frac{\langle \cos 2\varphi \rangle}{\langle \cos \varphi \rangle}$   as a function of rapidity difference Y
at NLL accuracy compared with a fixed order treatment. Right panel: Variation of the ratio $\frac{Y_{eff}}{Y}$ as a function of jet momentum ${\bf k}_{J,2}$ for fixed $ {\bf k}_{J,1}=35\,\mbox{GeV}$ for $Y=8$ and $s=7\,$TeV
at leading logarithmic (blue) and next-to-leading logarithmic (brown) accuracy
}
\label{asymmetric21+y}
\end{figure}

Up to now we discussed production of jets in the symmetric configuration. From theoretical point of view the Monte Carlo simulations suffer in this case from instabilities which makes difficult the comparison of theoretical results based on BFKL method with the fixed order  calculation. Such comparison of different theoretical predictions can be made in the case of jet production in the asymmetric configuration, in which two jets have very different transverse momenta.  
\begin{figure}[h]
\includegraphics[width=6.2cm]{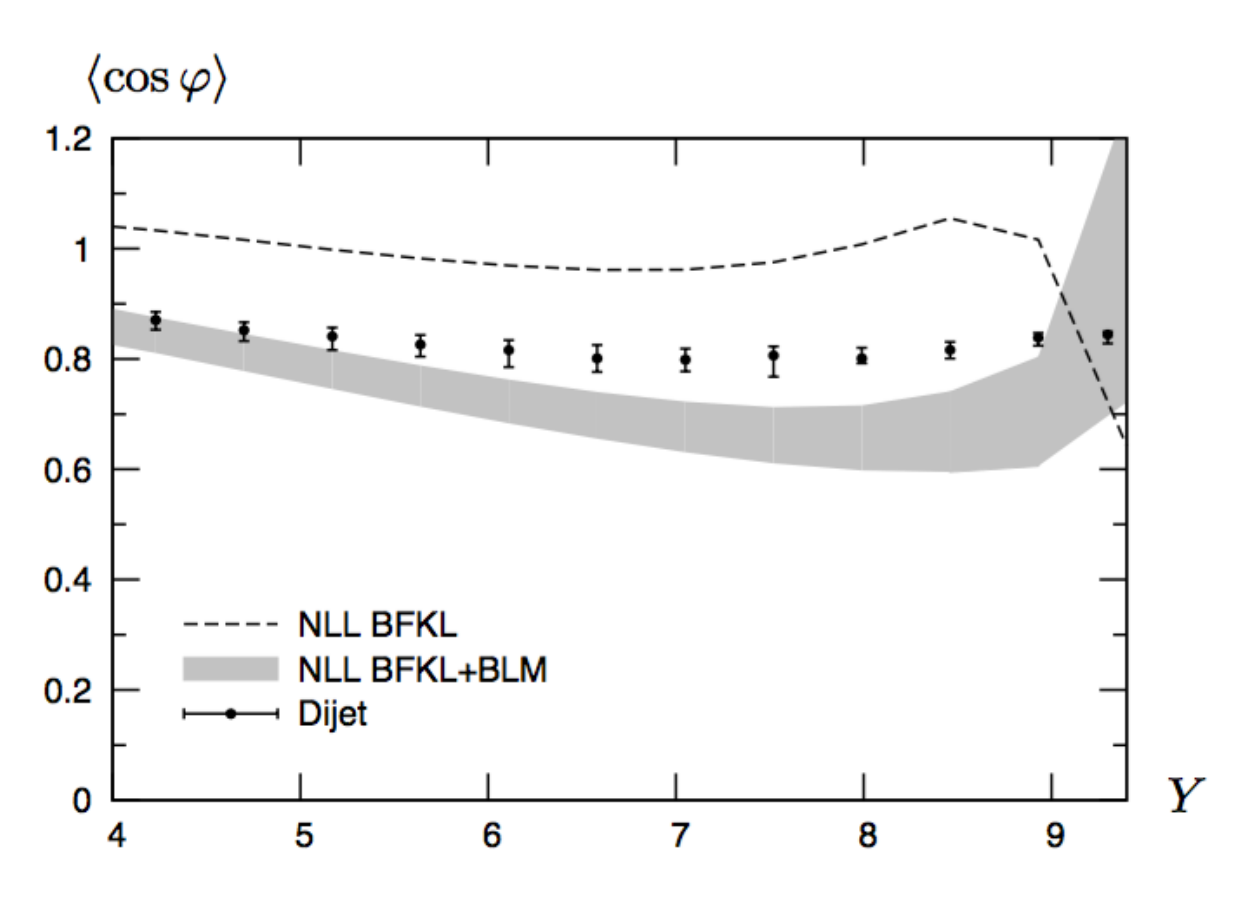}
\includegraphics[width=6.2cm]{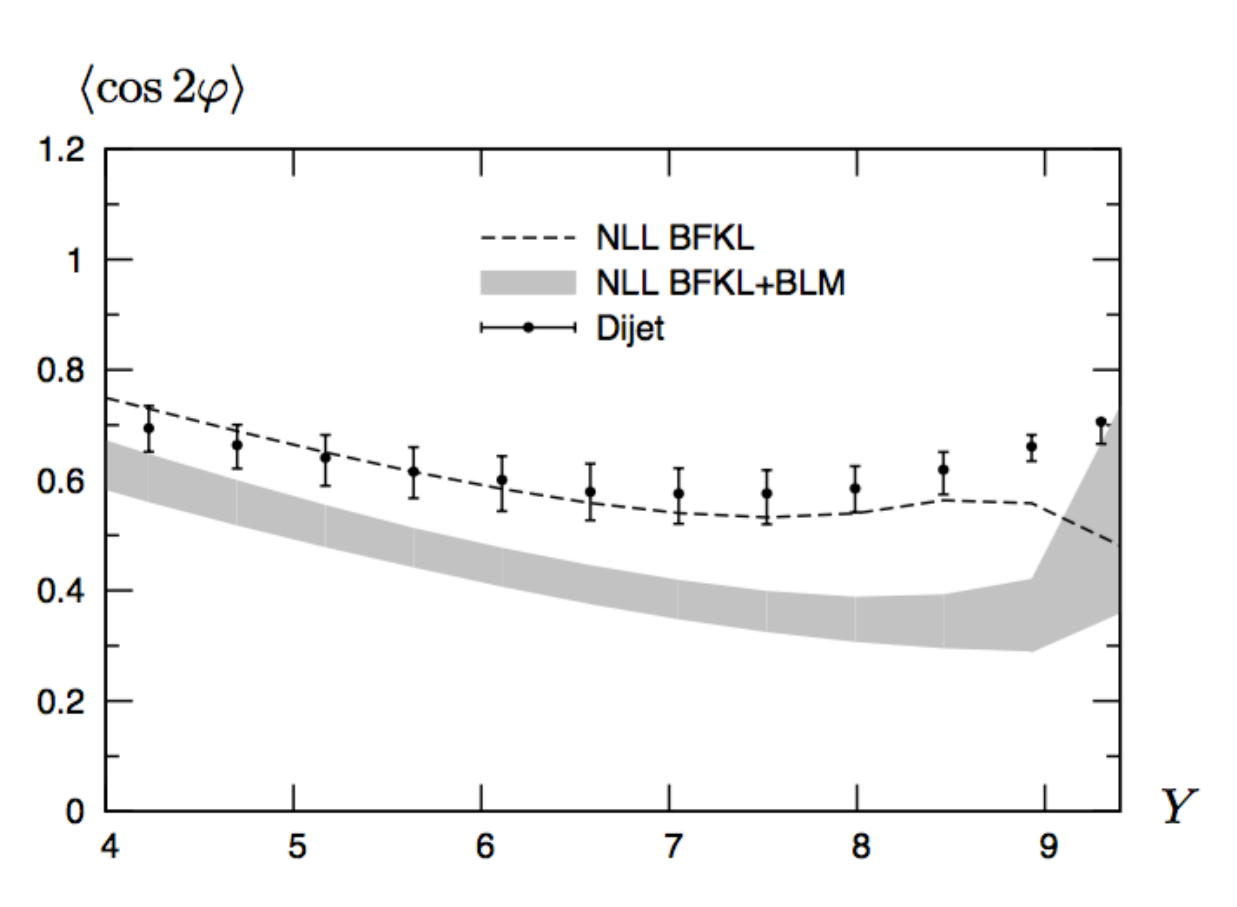}
 \caption{  
 Asymmetric configuration. Variation of $\langle \cos \varphi \rangle$ and $\langle \cos 2\varphi \rangle$ as a function of rapidity difference Y
at NLL accuracy compared with a fixed order treatment.
 }
\label{asymmetric}
\end{figure}
In the Fig. \ref{asymmetric} and in the left panel of Fig. \ref{asymmetric21+y} we present our theoretical predictions for decorrelation coefficients and their ratio confronted with the result of the fixed order  calculation of Ref. \cite{Aurenche:2008dn}.   It seems that specially in the case of the ratio $\frac{\cos 2\varphi}{\cos \varphi}$ a measurement of this observable 
could discriminate between two different  mechanisms.
Unfortunately, for now  experimental measurements in such asymmetric configurations, although very desirable,
 are not available.

The important drawback of the BFKL method is the fact that it does not respect exact energy-momentum conservation. This fact can lead to sizable numerically effects, although formally it represents a non-leading correction.   In the Ref. \cite{Ducloue:2014koa} we studied the violation of energy-momentum conservation for  asymmetric configuration using the 
method proposed by Del Duca nd Schmidt in 
\cite{DelDuca:1994ng}. In consist in introduction of the effective rapidity $ Y_{\rm eff} $ defined as
\begin{equation}
  Y_{\rm eff} \equiv\ Y \frac{\mathcal{C}_0^{2\to3}}{\mathcal{C}_0^{{\rm BFKL},\mathcal{O}(\alpha_s^3)}} 
  \label{eq:Cm_e-m_cons}
\end{equation}
in \cite{Ducloue:2014koa},
where $\mathcal{C}_0^{2\to3}$ is the amplitude for $2 \to 3$ partonic process (contributing to the cross section) calculated up to  $\mathcal{O}(\alpha_s^3)$  accuracy without  any approximations and   
$\mathcal{C}_0^{{\rm BFKL},\mathcal{O}(\alpha_s^3)}$
is the amplitude of the same process obtained within BFKL method.  If the violation of energy-momentum conservation is not numerically important the ratio $\frac{ Y_{\rm eff} }{Y}$ should take values close to one. In the right panel of Fig. \ref{asymmetric21+y} we show our result for the ratio $\frac{ Y_{\rm eff} }{Y}$ 
estimated by taking into account NLO BFKL corrections to the jet production vertex . We see that for very asymmetric jet momenta this ratio takes values close to 1, which justifies our conclusion that
the predictions obtained for production of jets in asymmetric configuration should not be affected by  violation of energy-momentum conservation.


%
%
%
%
%
%
%

\vspace*{+0.2cm}
This work is partly supported by grant No 2015/17/B/ST2/01838 of the National Science Center in Poland, by the French grant ANR PARTONS (Grant No. ANR-12-MONU-0008-01), by the Academy of Finland, project 273464, by the COPIN-IN2P3 agreement, by the Labex P2IO and by the Polish-French collaboration agreement Polonium.


\begin{thebibliography}{42}



\bibitem{Fadin:1975cbKuraev:1976geKuraev:1977fsBalitsky:1978ic}
V.S. Fadin, E.~Kuraev, L.~Lipatov, Phys. Lett. \textbf{B60}, 50 (1975);
%
Sov. Phys. JETP \textbf{44}, 443 (1976);
%
E.~Kuraev, L.~Lipatov, V.S. Fadin, Sov. Phys. JETP \textbf{45}, 199 (1977);
%
I.~Balitsky, L.~Lipatov, Sov. J. Nucl. Phys. \textbf{28}, 822 (1978)


\bibitem{Mueller:1986ey}
A.H. Mueller, H.~Navelet, Nucl. Phys. \textbf{B282}, 727 (1987)


\bibitem{Fadin:1998py}
V.S. Fadin, L.N. Lipatov, Phys. Lett. \textbf{B429}, 127 (1998),
  \texttt{hep-ph/9802290}

\bibitem{Ciafaloni:1998gs}
M.~Ciafaloni, G.~Camici, Phys. Lett. \textbf{B430}, 349 (1998),
  \texttt{hep-ph/9803389}

\bibitem{Bartels:2001ge}
J.~Bartels, D.~Colferai, G.P. Vacca, Eur. Phys. J. \textbf{C24}, 83 (2002),
  \texttt{hep-ph/0112283}

\bibitem{Bartels:2002yj}
J.~Bartels, D.~Colferai, G.P. Vacca, Eur. Phys. J. \textbf{C29}, 235 (2003),
  \texttt{hep-ph/0206290}
  
  
  \bibitem{Caporale:2011cc}
  F.~Caporale, D.~Y.~Ivanov, B.~Murdaca, A.~Papa and A.~Perri,
  JHEP {\bf 1202} (2012) 101, \texttt{hep-ph/1112.3752}
 
 \bibitem{Khachatryan:2016udy}
  V.~Khachatryan {\it et al.} [CMS Collaboration],
  JHEP {\bf 1608} (2016) 139, \texttt{hep-ex/1601.06713}
 
 

\bibitem{Colferai:2010wu}
D.~Colferai, F.~Schwennsen, L.~Szymanowski, S.~Wallon, JHEP \textbf{1012}, 026
  (2010), \texttt{1002.1365}

\bibitem{Ducloue:2013hia}
B.~Duclou\'e, L.~Szymanowski, S.~Wallon, JHEP \textbf{1305}, 096 (2013),
  \texttt{1302.7012}
  
  
\bibitem{Ducloue:2013bva}
B.~Duclou\'e, L.~Szymanowski, S.~Wallon, Phys. Rev. Lett. \textbf{112}, 082003
  (2014), \texttt{1309.3229}


\bibitem{Brodsky:1982gc}
S.J. Brodsky, G.P. Lepage, P.B. Mackenzie, Phys. Rev. \textbf{D28}, 228 (1983)

\bibitem{Aurenche:2008dn}
P.~Aurenche, R.~Basu, M.~Fontannaz, Eur. Phys. J. \textbf{C57}, 681 (2008),
  \texttt{0807.2133}

\bibitem{Ducloue:2014koa}
B.~Duclou\'e, L.~Szymanowski, S.~Wallon, Phys. Lett. \textbf{B738}, 311 (2014),
  \texttt{1407.6593}



\bibitem{DelDuca:1994ng}
V.~Del~Duca, C.R. Schmidt, Phys. Rev. \textbf{D51}, 2150 (1995),
  \texttt{hep-ph/9407359}





%
%
%
%
%
%
%
%
%
%
%
%
%
%
%
%
%
%
%
%
%
%
%
%
%
%
%
%
%
%
%
%
  
  



\end{thebibliography}
\end{document}